\documentclass[11pt,a4paper]{article}
\usepackage{amssymb}
\usepackage[dvips]{graphicx}
\usepackage{bm}
\usepackage{amsmath, textcomp}
\def\sB{\stackrel{\frown}{\bm{\Box}}}
\def\uB{\stackrel{\frown}{\Box}}

\numberwithin{equation}{section}

\begin{document}

\begin{center}
\vspace{1cm} {\Large\bf Supergraph calculation of one-loop divergences}\\
\medskip
{\Large\bf in  higher-derivative $6D$ SYM theory} \vspace{1.0cm}

 {\bf
 I.L. Buchbinder\footnote{joseph@tspu.edu.ru }$^{\,a,b,c}$,
 E.A. Ivanov\footnote{eivanov@theor.jinr.ru}$^{\,c}$,
 B.S. Merzlikin\footnote{merzlikin@tspu.edu.ru}$^{\,d,a,c}$,
 K.V. Stepanyantz\footnote{stepan@m9com.ru}$^{\,e,c}$
 }
\vspace{0.4cm}

{\it $^a$ Department of Theoretical Physics, Tomsk State Pedagogical
 University,\\ 634061, Tomsk,  Russia \\ \vskip 0.15cm
 $^b$ National Research Tomsk State University, 634050, Tomsk, Russia \\ \vskip 0.1cm
 $^c$ Bogoliubov Laboratory of Theoretical Physics, JINR, 141980 Dubna, Moscow region,
 Russia \\ \vskip 0.1cm
 $^d$ Tomsk State University of Control Systems and Radioelectronics, 634050 Tomsk,
 Russia\\ \vskip 0.1cm
 $^e$ Department of Theoretical Physics, Moscow State University,
119991, Moscow, Russia }
\end{center}
\vspace{0.4cm}

\begin{abstract}
We apply the harmonic superspace approach for calculating the divergent part
of the one-loop effective action of renormalizable $6D$, ${\cal N}=(1,0)$ supersymmetric higher-derivative gauge theory
with a dimensionless coupling constant.
Our consideration uses the background superfield method allowing
to carry out the analysis of the effective action in a manifestly gauge
covariant and ${\cal N}=(1,0)$ supersymmetric way. We exploit the regularization by dimensional reduction in which
the divergences are absorbed into a renormalization of the coupling
constant. Having the expression for the one-loop divergences, we calculate the relevant
$\beta$-function. Its sign is specified by the overall sign of the classical action
which in higher-derivative theories is not fixed {\it a priori}. The result agrees
 with the earlier calculations in the component approach. The superfield calculation is
 simpler and provides possibilities for various generalizations.
\end{abstract}

\setcounter{footnote}{0}

\section{Introduction}
\hspace*{\parindent}

It is well known that field theories with standard kinetic terms and
standard interactions are not renormalizable in higher dimensions
because of the dimensionful coupling constants and the too slow decrease of
propagators at large momenta. Supersymmetry  is sometimes capable to improve the ultraviolet
behavior in the lowest loops
\cite{Fradkin:1982kf,Marcus:1983bd,Marcus:1984ei,Howe:1983jm}, but
does not lead to the renormalizability \cite{Howe:1983jm} even
in theories with the maximally extended supersymmetry
\cite{Howe:2002ui,Bossard:2009sy,Bossard:2009mn,Bork:2015zaa}.

A way to obtain a higher-dimensional renormalizable theory is to
allow the action to include terms with higher derivatives. The
study of the higher derivative theories at the classical and quantum
levels has a long history that apparently begins with the seminal work
by Pais and Uhlenbeck \cite{Pais:1950za}.  Although the higher-derivative theories
suffer from the ghost states in
the spectrum, they still attract much attention and are widely used in
gravity, cosmology and quantum field theory (see, e.g.,
\cite{Starobinsky:1980te,Hawking:1985gh,Stelle:1976gc,Fradkin:1981iu,Buchbinder:1992rb}).
Some of the recent applications, as well as the discussion of ways to evade the problem of ghosts, can be found in
Ref. \cite{Grinstein:2008qq,Beccaria:2015ypa,Osborn:2016bev,Kuzenko:2017jdy,Castellanos:2018dub,Anselmi:2020tqo}
and reviews \cite{Smilga:2016dpe,Casarin:2017xez} (and the references
therein). So the higher-derivative models are considered to be very interesting
and deserving the study from different points of view.

An important example of application of the higher-derivative theories is
the regularization by higher-order covariant derivatives
\cite{Slavnov:1971aw,Slavnov:1972sq}.
This regularization is self-consistent and, for supersymmetric
theories, can be formulated in a manifestly supersymmetric way
\cite{Krivoshchekov:1978xg,West:1985jx,Buchbinder:2014wra,Buchbinder:2015eva}
consistent with the non-renormalization theorems (see, e.g.,
\cite{Gates:1983nr,West:1990tg,Buchbinder:1998qv}). When applied for investigating quantum
corrections in supersymmetric field theories,  the higher-derivative regularization allowed
to reveal some interesting features of them (see, e.g.,
\cite{Stepanyantz:2019lyo,Stepanyantz:2019lfm} and references
therein). This is one more argument why it is useful to study  the quantum
corrections in higher-derivative supersymmetric theories in various
dimensions.

In this paper we consider the six-dimensional higher-derivative supersymmetric gauge theory proposed in Ref. \cite{Ivanov:2005qf} and calculate the divergent part
of the one-loop effective action, using the regularization by dimensional reduction. This theory describes the
following set of the interacting $6D$ fields: the vector field, the
Weyl spinor field, and three real scalar fields, all being in the
adjoint representation of the gauge group. In the gauge field sector the action
starts with the term containing four derivatives

\begin{equation}\label{Vector_Part}
- \frac{1}{g_0^2} \mbox{tr} \int d^6x\, (\nabla^M F_{MN})^2,
\end{equation}

\noindent where $F_{MN}$ is the standard Yang-Mills strength.
This implies that the coupling constant $g_0$ is dimensionless.
The manifestly supersymmetric formulation of the theory in $6D$, ${\cal N}=(1,0)$ harmonic superspace and the full off-shell component form of the action
in the Wess-Zumino gauge were earlier given in Ref. \cite{Ivanov:2005qf}\footnote{The radical difference of such higher-derivative theory from the theories regularized by higher-order
derivatives is that the actions of the latter theories involve higher-derivative terms as corrections to the standard kinetic terms with canonical numbers of derivatives (two for bosons and
one for fermions), while in the case under consideration no canonical kinetic terms are present from the very beginning.}.

In the case under consideration, the issue of renormalizabity was
studied in \cite{Ivanov:2005qf} (see also \cite{Casarin:2019aqw}) in
a component formulation and in \cite{Buchbinder:2020ovf} in a
superfield formalism \cite{Buchbinder:2020ovf}. The actual
calculations definitely show that the theory is renormalizable at
one loop.\footnote{The theory under consideration possesses a chiral
anomaly \cite{Smilga:2006ax} which, in principle, can violate the
renormalizability. However it is known that the anomaly does not
affect the form of one-loop divergences.} The one-loop divergences, renormalization of the coupling constant
and the corresponding beta-function in this theory were
calculated in Refs. \cite{Ivanov:2005qf,Casarin:2019aqw} by two
different methods based on the component formulations.

The study of various aspects of the four-dimensional supersymmetric
quantum field theories (see, e.g., the monographs
\cite{Gates:1983nr,West:1990tg,Buchbinder:1998qv}) provided an
evidence that the most attractive and elegant way of investigating
their quantum properties is by using superfield methods. The
superfield formulation of $6D$ supersymmetric theories was
constructed in Refs.
\cite{Howe:1985ar,Zupnik:1986da,Bossard:2015dva} in  terms of $6D$,
${\cal N}=(1,0)$ harmonic superspace which is quite similar to its
$4D$, ${\cal N}=2$ prototype
\cite{Galperin:1985ec,Galperin:1984av,Galperin:2001uw}. The main
advantage of such a formulation is the possibility to keep manifest
${\cal N}=(1,0)$ supersymmetry at all steps  of quantum
calculations. In our recent papers
\cite{Buchbinder:2016gmc,Buchbinder:2016url,Buchbinder:2017ozh,Buchbinder:2017gbs,Buchbinder:2017xjb,Buchbinder:2018bhs,Buchbinder:2018lbd}
we  developed the harmonic superfield approach for calculating the
lowest off-shell quantum corrections in various $6D$, ${\cal
N}=(1,0)$ and ${\cal N}=(1,1)$ supersymmetric theories. In the
present paper we apply this superfield technique for studying the
one-loop effective action in $6D$, ${\cal N}=(1,0)$
higher-derivative gauge theory of Ref. \cite{Ivanov:2005qf}.

The paper is organized as follows. In Sect.
\ref{Section_Harmonic_Superspace} we collect the basic notions of
$6D$, ${\cal N}=(1,0)$  harmonic superspace and the formulation of
the model under consideration within its framework. Sect.
\ref{Section_Quantization} presents the manifestly supersymmetric
and gauge covariant quantization of this theory and the construction
of the corresponding effective action. In Sect.
\ref{Section_One-Loop} we compute the one-loop divergences by a
direct calculation of the harmonic supergraphs and find the
$\beta$-function. The results and some further problems are
summarized in Conclusion. The technical details of the calculation
are contained in Appendices A and B.

\section{Harmonic superspace formulation of  $6D$, ${\cal N}=(1,0)$ higher-derivative SYM theory}
\hspace*{\parindent}\label{Section_Harmonic_Superspace}

The harmonic superspace technique is most convenient for
formulating theories with  $6D$, ${\cal N}=(1,0)$
supersymmetry as it suggests the manifestly supersymmetric and
gauge invariant scheme of their quantization.

In our notation the coordinates of $6D$ Minkowski
space and the ${\cal N}=(1,0)$ Grassmann coordinates are denoted by
$x^M$ and $\theta^{ai}$, with $M=0,\ldots,5$,\ $a=1,\ldots, 4$, and
$i=1,2$. The coordinates $(x^M,\theta^{ai}, u_i^\pm)$ of the
harmonic superspace  include in addition the harmonic variables $u_i^\pm$
which obey the constraints $u^{+i} u_i^- = 1$, $u_i^- \equiv (u^{+i})^*$. Having these coordinates at hand, one can construct the harmonic
derivatives

\begin{equation}
D^{++} = u^{+i} \frac{\partial}{\partial u^{-i}};\qquad D^{--} = u^{-i} \frac{\partial}{\partial u^{+i}};\qquad
D^0 = u^{+i} \frac{\partial}{\partial u^{+i}} - u^{-i} \frac{\partial}{\partial u^{-i}},
\end{equation}

\noindent
which generate an $SU(2)$ algebra. The harmonic superspace contains an analytic subspace
closed under the $6D, \,{\cal N}=(1,0)$ supersymmetry transformations. It is parametrized by the coordinates

\begin{equation}
x^M_A = x^M + \frac{i}{2}\theta^{-}\gamma^M \theta^+; \qquad \theta^{\pm a} = u^\pm_i \theta^{ai},
\end{equation}

\noindent
where $\gamma^M$ are $6D$  $\gamma$-matrices.

We also introduce the harmonic spinor covariant derivatives

\begin{equation}
D^+_a = u^{+}_i D_{a}^i;\qquad D^-_a = u^{-}_i D_{a}^i\,,
\end{equation}

\noindent
the only non-zero anticommutation relation among which being $\{D^+_a, D^{-}_b\} = i(\gamma^M)_{ab}\partial_M$. Due to the anticommutativity of the derivatives $D^+_a$ any product of four such derivatives (defined with respect to
the same harmonic variable $u$) is reduced to the expression

\begin{equation}
(D^+)^4 = -\frac{1}{24}\varepsilon^{abcd} D_a^+ D_b^+ D_c^+ D_d^+.
\end{equation}

In this paper we adopt the following convention for the superspace integration measures needed for constructing
the ${\cal N}=(1,0)$ invariant actions:

\begin{equation}
\int d\zeta^{(-4)} \equiv \int d^6x\, d^4\theta^+;\qquad \int d^{14}z \equiv \int d^6x\,d^8\theta = \int d^6x\,d^4\theta^{+} (D^+)^4.
\end{equation}

In the harmonic superspace formulation the gauge field is carried by the superfield $V^{++}(z,u) = V^{++A} t^A$ which obeys the
analyticity condition

\begin{equation}
D^+_a V^{++} = 0
\end{equation}

\noindent
and is real with respect to a generalized conjugation denoted by a tilde, $\widetilde{V^{++}} = V^{++}$.
In this paper we use the Hermitian generators $t^A$ which are normalized by the conditions $\mbox{tr}(t^A t^B) = \delta^{AB}/2$.
From the gauge superfield $V^{++}$ one can construct a non-analytic superfield

\begin{equation}\label{V--_Definition}
V^{--}(z,u) \equiv \sum\limits_{n=1}^\infty (-i)^{n+1} \int du_1 du_2 \ldots du_n \frac{V^{++}(z,u_1) V^{++}(z,u_2)
\ldots V^{++}(z,u_n)}{(u^+u_1^+)(u_1^+ u_2^+) \ldots (u_n^+ u^+)},
\end{equation}

\noindent
and, further, the harmonic gauge superfield strength

\begin{equation}\label{F++}
F^{++} \equiv (D^+)^4 V^{--}.
\end{equation}

\noindent
It is evidently analytic. Moreover, it satisfies the off-shell condition

\begin{equation}\label{Equation_For_F}
\nabla^{++} F^{++} \equiv D^{++} F^{++} + i [V^{++}, F^{++}] = 0\,,
\end{equation}

\noindent
which is a consequence of the harmonic flatness condition

\begin{equation}\label{Flatness}
D^{++} V^{--} - D^{--}V^{++} + i [V^{++}, V^{--}] = 0\,.
\end{equation}

\noindent
The latter can be considered as a definition of $V^{--}$.

The gauge transformations in the harmonic superspace are parametrized by the real (with respect to the tilde-conjugation)
analytic superfield $\lambda = \lambda^A t^A$:

\begin{eqnarray}\label{Gauge_Invariance}
V^{\pm\pm} \to e^{i\lambda} V^{\pm\pm} e^{-i\lambda} - i e^{i\lambda} D^{\pm\pm} e^{-i\lambda},\qquad\quad
F^{++} \to e^{i\lambda} F^{++} e^{-i\lambda}\,.\qquad \nonumber\\
\end{eqnarray}

The $6D$, ${\cal N}=(1,0)$ supersymmetric generalization of the usual $6D$ Yang--Mills theory
in the harmonic superspace formulation is given by the action \cite{Zupnik:1987vm}

\begin{equation}\label{Usual_SYM}
S_{\mbox{\scriptsize SYM}} = \frac{1}{f_0^2} \sum\limits_{n=2}^\infty \frac{(-i)^{n}}{n} \mbox{tr} \int d^{14}z\, du_1 \ldots du_n\,
\frac{V^{++}(z,u_1)\ldots V^{++}(z,u_n)}{(u_1^+ u_2^+) \ldots (u_n^+ u_1^+)},
\end{equation}

\noindent with the coupling constant $f_0$ having the
dimension of the inverse mass. It is clear that this  theory is not renormalizable by power counting. The one-loop divergences for this theory have been calculated in Ref. \cite{Buchbinder:2017ozh}.

In the present paper we will consider a different theory
which contains the higher (four) derivatives. Unlike the second-order derivative SYM theory with action (2.12), the
higher-derivative theory we are considering is characterized by a dimensionless coupling
constant. Such a theory was formulated in harmonic $6D$ superspace
in \cite{Ivanov:2005qf}. It is described by the following manifestly
gauge invariant and ${\cal N}=(1,0)$ supersymmetric action

\begin{equation}\label{Action}
S = \pm \frac{1}{2g_0^2}\, \mbox{tr} \int d\zeta^{(-4)} du\,
(F^{++})^2 = \pm \frac{1}{4g_0^2} \int d\zeta^{(-4)} du
\big(F^{++A}\big)^2\,,
\end{equation}

\noindent
where the analytic harmonic superfield strength $F^{++}$ is defined in Eq. (\ref{F++}).
The aim of our paper is to investigate the one-loop divergences
for the theory with the action (\ref{Action}).

The sign of the action (\ref{Action}) deserves some comments.  In
conventional field theory models without higher derivatives the
overall sign is fixed by the requirement that the energy is
positive. In the higher-derivative theories the energy is not
positively defined in general. This means, that there are no actual reasons to
fix an overall sign of the action in such theories. This is why we
cannot fix the sign of the action (\ref{Action}) \footnote{This
point was also noted in \cite{Casarin:2019aqw}.}. Note that in Ref.
\cite{Ivanov:2005qf} there was chosen the sign minus (corresponding to the sign in (\ref{Vector_Part})) since it  gives
rise to the correct sign in front of the component kinetic
term of the triplet of scalar fields entering the gauge ${\cal
N}=(1,0)$ multiplet (the former auxiliary fields). However, this
does not imply the positivity of energy for all component
fields which involve higher-derivative ghosts for any sign. In Ref. \cite{Casarin:2019aqw} there was chosen the sign ``plus''.
In order to have a freedom to compare our results with those obtained in Refs. \cite{Ivanov:2005qf,Casarin:2019aqw},
we prefer not to fix the overall sign of the action.

\section{Background field quantization in harmonic superspace}
\hspace*{\parindent}\label{Section_Quantization}

The harmonic superspace technique makes it possible to construct the manifestly ${\cal N}=(1,0)$ supersymmetric quantization procedure.
It is also convenient to use the background superfield method \cite{DeWitt:1965jb,Abbott:1980hw,Abbott:1981ke}
which provides a manifestly gauge invariant effective action. In $6D$, ${\cal N}=(1,0)$ harmonic superspace
it is formulated similarly to the $4D$, ${\cal N}=2$ case treated
in \cite{Buchbinder:1997ya,Buchbinder:2001wy}. In particular, the background-quantum splitting is linear,

\begin{equation}\label{Splitting}
V^{++} = \bm{V}^{++} + v^{++},
\end{equation}

\noindent
where $\bm{V}^{++}$ and $v^{++}$ are the background and quantum gauge superfields, respectively.
After the substitution of (\ref{Splitting}) in the action (\ref{Action}) the gauge invariance (\ref{Gauge_Invariance})
amounts to the two types of transformations. The background gauge invariance

\begin{equation}\label{Background_Symmetry}
\bm{V}^{++} \to e^{i\lambda} \bm{V}^{++} e^{-i\lambda} - i e^{i\lambda} D^{++} e^{-i\lambda};\qquad v^{++} \to e^{i\lambda} v^{++} e^{-i\lambda}
\end{equation}

\noindent
remains a manifest symmetry of the effective action, while the quantum gauge invariance

\begin{equation}
\bm{V}^{++} \to \bm{V}^{++};\qquad v^{++} \to e^{i\lambda} (v^{++} + \bm{V}^{++}) e^{-i\lambda} -\bm{V}^{++} - i e^{i\lambda} D^{++} e^{-i\lambda}
\end{equation}

\noindent
is broken by the gauge-fixing procedure down to the invariance under the BRST transformations.
It is assumed that the gauge-fixing term should be chosen invariant under the transformations (\ref{Background_Symmetry}).
The harmonic superspace analog of the $\xi$-gauge is then given by the action

\begin{equation}\label{Gauge-Fixing_Action}
S_{\mbox{\scriptsize gf}} = \mp \frac{1}{2 g_0^2 \xi_0}\, \mbox{tr} \int d^{14}z\, du_1\, du_2\, \frac{(u_1^- u_2^-)}{(u_1^+ u_2^+)^3}\, e^{i\bm{b}_1} e^{-i\bm{b}_2} (\bm{\nabla}_2^{++} v_{2}^{++})\, e^{i\bm{b}_2} e^{- i\bm{b}_1} \sB_1 (\bm{\nabla}_1^{++} v_{1}^{++}),
\end{equation}

\noindent
where the operator

\begin{equation}\label{Box_Definition}
\sB\ \equiv \frac{1}{2} (D^+)^4 (\bm{\nabla}^{--})^2
\end{equation}

\noindent is reduced to the covariant analog of the d'Alambertian
operator, when acting on the analytic
superfields \footnote{In principle, in the framework of the background field method we could use any appropriate
gauge preserving the background gauge
invariance. However, it is technically convenient to choose the action $S_{\mbox{\tiny gf}}$ to be of
the same degree in derivatives as the classical
action.}. The background covariant derivatives are defined as

\begin{equation}\label{Background_Covariant_Derivatives}
\bm{\nabla}^{++} = D^{++} + i \bm{V}^{++}; \qquad \bm{\nabla}^{--} = D^{--} + i \bm{V}^{--}.
\end{equation}

\noindent
Evidently, if they act on a superfield in the adjoint representation (e.g., on $v^{++}$),
the gauge superfields should be expanded over the generators of the adjoint representation, so that

\begin{equation}
\bm{\nabla}^{\pm\pm} v^{++} = D^{\pm\pm} v^{++} + i [\bm{V}^{\pm\pm}, v^{++}].
\end{equation}

\noindent
The background bridge superfield $\bm{b}$ in Eq. (\ref{Gauge-Fixing_Action})
is related to the background superfields $\bm{V}^{++}$ and $\bm{V}^{--}$ (constructed out of $\bm{V}^{++}$
by the equation similar to (\ref{V--_Definition})) via the relations

\begin{equation}\label{Background_Bridge}
\bm{V}^{++} = -i e^{i\bm{b}} D^{++} e^{-i\bm{b}};\qquad \bm{V}^{--} = -i e^{i\bm{b}} D^{--} e^{-i\bm{b}}.
\end{equation}

\noindent
The subscripts $1$ and $2$ in Eq. (\ref{Gauge-Fixing_Action}) refer to the harmonic variables $u_1$ or $u_2$
present in $S_{\rm gf}$. Similar notation will be used below.

The action for the Faddeev--Popov ghosts corresponding to the gauge-fixing action (\ref{Gauge-Fixing_Action})
is obtained in the standard way (see, e.g., \cite{Buchbinder:1997ya}) and is given by

\begin{equation}
S_{\mbox{\scriptsize FP}} = \mbox{tr} \int d\zeta^{(-4)} du\, b \bm{\nabla}^{++}\Big(\bm{\nabla}^{++} c + i[v^{++},c]\Big),
\end{equation}

\noindent
where the anticommuting analytic superfields $b$ and $c$ stand for the Faddeev--Popov antighosts and ghosts, respectively.
However, the presence of the operator $\sB$ in Eq. (\ref{Gauge-Fixing_Action}) change the form of the Nielsen--Kallosh determinant.
Namely, in the case under consideration it can be written in the form

\begin{equation}
\Delta_{\mbox{\scriptsize NK}} = \mbox{Det}^{-1/2}(\bm{\nabla}^{++})^2\, \mbox{Det} \sB\  = \int D\varphi\, D\chi^{(+4)} D\sigma\, \exp\left(iS_{\mbox{\scriptsize NK}}\right),
\end{equation}

\noindent
where the set of Nielsen--Kallosh ghosts involves the commuting analytic superfield $\varphi$
together with the Grassmann-odd analytic superfields $\chi^{(+4)}$ and $\sigma$, all being in the adjoint representation of the gauge group.
The action for these ghosts reads

\begin{equation}
S_{\mbox{\scriptsize NK}} = \mbox{tr} \int d\zeta^{(-4)} du\,\Big(-\frac{1}{2} (\bm{\nabla}^{++}\varphi)^2 + \chi^{(+4)}\sB \sigma\Big).
\end{equation}

\noindent
Then the generating functional for the considered theory can finally be written in the form

\begin{equation}
Z[\,\mbox{Sources}, \bm{V}^{++}] = \int Dv^{++} Db\, Dc\, D\varphi\, D\chi^{(+4)} D\sigma\, \exp\Big\{i\Big(S + S_{\mbox{\scriptsize gf}} + S_{\mbox{\scriptsize FP}} + S_{\mbox{\scriptsize NK}} + S_{\mbox{\scriptsize sources}}\Big)\Big\}.
\end{equation}

\noindent
The source term is defined as

\begin{equation}
S_{\mbox{\scriptsize sources}} = \int d\zeta^{(-4)} du\, J^{++A} v^{++A} + \ldots,
\end{equation}

\noindent
where dots denote terms corresponding to various ghost superfields.
The effective action $\Gamma[\,\mbox{Fields}, \bm{V}^{++}]$ is defined as the Legendre transform
of the generating functional for the connected Green functions $W\equiv -i\ln Z$.
Setting all quantum fields equal to zero, we obtain the manifestly gauge invariant
action $\Gamma[\bm{V}^{++}]\equiv \Gamma[\,\mbox{Fields}\to 0, \bm{V}^{++}]$.

Like in the case of standard $6D, \, {\cal N}=(1,0)$ SYM theory \cite{Buchbinder:2016url, Buchbinder:2017ozh},
there are two ways to calculate divergent terms in this action. One of them is based on the superfield
proper-time technique and preserves manifest gauge
invariance at al steps. Another goes through the direct calculation of the relevant Feynman supergraphs
with invoking gauge invariance at the final stage. Here we employ the second method, leaving the first one for
the future study.

\section{Calculation of the divergent supergraphs}
\hspace*{\parindent}\label{Section_One-Loop}

\begin{figure}[h]
\begin{picture}(0,2)
\put(0,1.8){(1)}
\put(0.1,0.2){\includegraphics[scale=0.45]{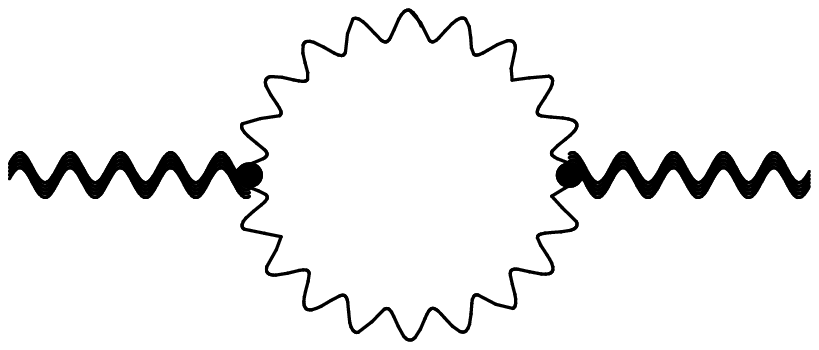}}
\put(4.5,1.8){(2)}
\put(4.6,0){\includegraphics[scale=0.45]{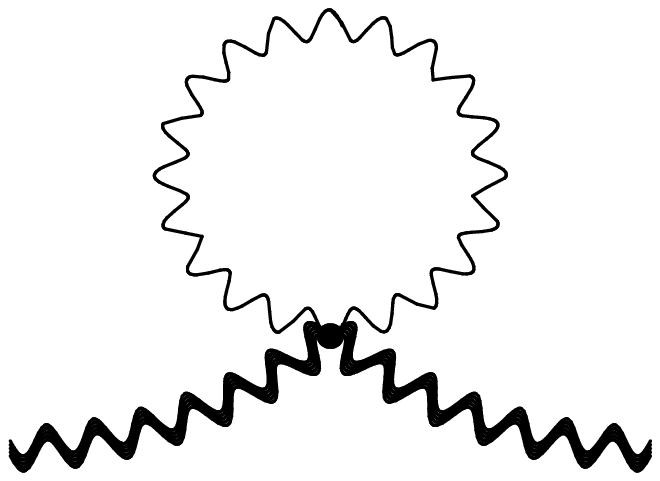}}
\put(8.3,1.8){(3)}
\put(8.4,0.2){\includegraphics[scale=0.45]{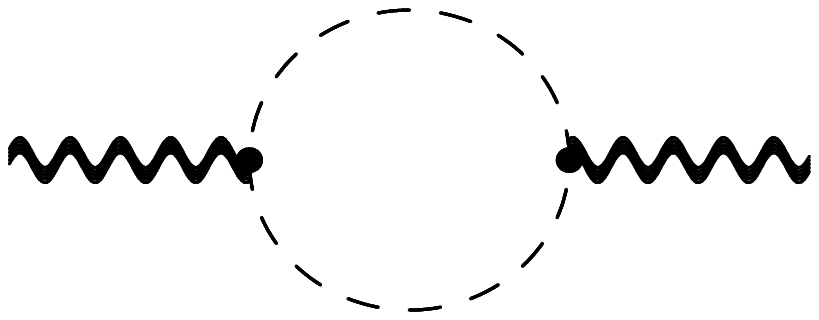}}
\put(12.85,1.8){(4)}
\put(12.9,0){\includegraphics[scale=0.45]{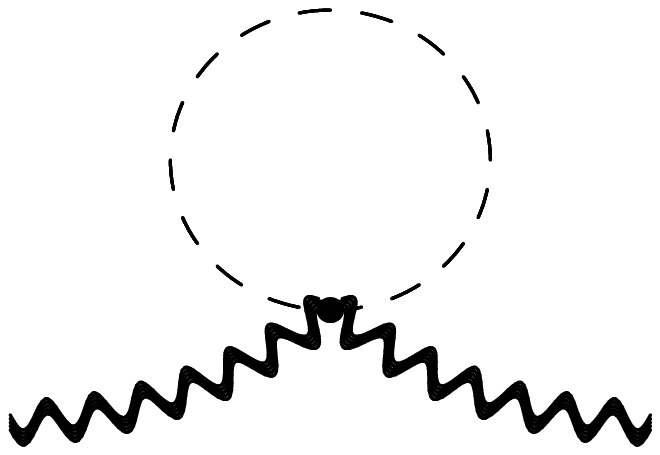}}
\end{picture}
\caption{Supergraphs which allow to calculate the divergent part of the one-loop effective action for the model (\ref{Action}).
The dashed lines stand for all ghost superfields present in the theory.}\label{Figure_Diagrams}
\end{figure}

The one-loop contribution to the two-point Green function of the background gauge superfield for the considered model is given
by the supergraphs depicted on Fig. \ref{Figure_Diagrams}. We will calculate it in the minimal gauge $\xi=1$,
where $\xi$ is the renormalized gauge-fixing parameter. The external wavy lines in the supergraphs on
Fig. \ref{Figure_Diagrams} represent the background gauge superfield $\bm{V}^{++}$ (or to the background bridge $\bm{b}$).
The wavy internal lines correspond to the propagators of the quantum gauge superfield $v^{++}$. In the Feynman gauge
this propagator has the simplest form and is given by the expression

\begin{equation}
- \frac{\delta^2 Z_0}{\delta J_1^{++A} \delta J_2^{++B}}\bigg|_{J^{++}=0} = \pm 2i g_0^2\, \frac{1}{\partial^4} (D_2^+)^4 \delta^{14}(z_1-z_2)\, \delta^{(2,-2)}(u_1,u_2),
\end{equation}

\noindent
where

\begin{equation}
\delta^{14}(z_1-z_2) \equiv \delta^6(x_1-x_2)\, \delta^8(\theta_1-\theta_2)
\end{equation}

\noindent and $Z_0$ denotes the generating functional of the free
theory. The dashed lines denote propagators of the Faddeev--Popov
and Nielsen--Kallosh ghosts. All ghost contributions have already
been calculated earlier \cite{Buchbinder:2016url,Buchbinder:2017ozh}.
Actually, the ghost part of the generating functional for the theory under consideration and for the theory
(\ref{Usual_SYM}) differ only in the expression

\begin{equation}
\bigg(\int D\chi^{(+4)} D\sigma\, \exp\Big\{i\,\mbox{tr} \int d\zeta^{(-4)} du\,\chi^{(+4)} \sB \sigma\Big\}\bigg)^{1/2}.
\end{equation}

\noindent
However, according to \cite{Buchbinder:2017ozh} this expression cannot produce divergences.
This implies that the total ghost contribution (which include both Faddeev--Popov and Nielsen--Kallosh parts
and can be found by calculating the supergraphs (3) and (4) in Fig. \ref{Figure_Diagrams}) coincides with
the one for the theory (\ref{Usual_SYM}) and, according to \cite{Buchbinder:2017ozh}, is equal to

\begin{equation}\label{Ghost_Contribution}
\big(\Delta\Gamma^{(1)}_\infty\big)_{\mbox{\scriptsize ghost}} = \frac{C_2}{3\varepsilon (4\pi)^3}\, \mbox{tr} \int d\zeta^{(-4)} du\, (\bm{F}^{++})^2,
\end{equation}

\noindent
where it is assumed that the regularization by dimensional reduction is used, with $\varepsilon \equiv 6-D$. The constant $C_2$ is defined by the equation $f^{ACD} f^{BCD} = C_2 \delta^{AB}$, where the structure constants $f^{ABC}$ are given by the commutator of generators, $[t^A, t^B] = i f^{ABC} t^C$.

Thus, it remains to calculate only the one-loop divergences produced by the supergraphs containing a loop
of the gauge quantum superfield. These divergences are completely determined by the part of the total action
quadratic in the quantum gauge superfield (which does not contain other quantum superfields).
Obviously, such terms are present only in the classical action (\ref{Action}) and the gauge-fixing action (\ref{Gauge-Fixing_Action}).
In Appendix \ref{Appendix_Variation} we demonstrate that, in the Feynman gauge
$\xi_0=1$\footnote{When calculating the one-loop divergences, there is no difference
between choices $\xi_0=1$ and $\xi=1$.},

\begin{eqnarray}\label{Second_Variation}
&& S^{(2)} + S_{\mbox{\scriptsize gf}} = \pm \frac{1}{2g_0^2} \mbox{tr} \int d\zeta^{(-4)}\, du\, v^{++} \sB{\hspace*{-0.5mm}}^2 v^{++} \qquad \nonumber\\
&&\qquad\qquad\quad\ \ \mp \frac{i}{2g_0^2} \mbox{tr} \int d^{14}z\, du_1\, du_2\, \frac{1}{(u_1^+ u_2^+)^2} e^{i\bm{b}_1} e^{-i\bm{b}_2}  v_2^{++} e^{i\bm{b}_2} e^{-i\bm{b}_1} [\bm{F}_1^{++}, \bm{\nabla}^{--}_1 v^{++}_1].\qquad
\end{eqnarray}

\noindent
Starting from this expression we calculate the one-loop divergences
coming from the supergraphs which contain a loop of the quantum gauge superfield (see Appendix \ref{Appendix_Divergences}).
Namely, from the expression (\ref{Second_Variation}) we find the vertices in the supergraphs depicted on Fig. \ref{Figure_Diagrams}.
Next, we calculate the first two supergraphs in Fig. \ref{Figure_Diagrams}
and obtain the corresponding contribution to the divergent part of the two-point Green function of the background gauge superfield.
Finally, taking into account the background gauge invariance of the effective action,
we obtain the general result for all one-loop divergences coming from the supergraphs with a loop of the quantum gauge superfield
inside. The result is given by the expression

\begin{equation}\label{Gauge_Contribution}
\big(\Delta\Gamma^{(1)}_\infty\big)_{\mbox{\scriptsize gauge}} = -\frac{4C_2}{\varepsilon (4\pi)^3}\, \mbox{tr} \int d\zeta^{(-4)} du\, (\bm{F}^{++})^2.
\end{equation}

\noindent
Summing up Eqs. (\ref{Ghost_Contribution}) and (\ref{Gauge_Contribution}), we obtain that the divergent part
of the one-loop effective action for the theory (\ref{Action}) regularized by the dimensional reduction takes the form

\begin{equation}\label{Quantum_Correction}
\Delta\Gamma^{(1)}_\infty = \big(\Delta\Gamma^{(1)}_\infty\big)_{\mbox{\scriptsize gauge}} + \big(\Delta\Gamma^{(1)}_\infty\big)_{\mbox{\scriptsize ghost}} = - \frac{11 C_2}{3\varepsilon (4\pi)^3}\, \mbox{tr} \int d\zeta^{(-4)} du\, (\bm{F}^{++})^2.
\end{equation}

\noindent
This result exactly agrees with the ones obtained in Refs. \cite{Ivanov:2005qf,Casarin:2019aqw} starting from the
component formulation of the theory in the Wess--Zumino gauge.

Adding the expression (\ref{Quantum_Correction}) to the classical action (\ref{Action}) we can find quantum corrections
to the coupling constant $g_0$,

\begin{eqnarray}\label{Effective_Action}
&& \Gamma - S_{\mbox{\scriptsize gf}} = \Big(\pm \frac{1}{2g_0^2} - \frac{11 C_2}{3\varepsilon(4\pi)^3} \Big)\, \mbox{tr} \int d\zeta^{(-4)} du\, (\bm{F}^{++})^2  \qquad\nonumber\\
&&\qquad\qquad + \mbox{finite one-loop contributions} + \mbox{higher order corrections}.\qquad\vphantom{\Big(}
\end{eqnarray}

\noindent
From this expression we see that the renormalized coupling constant $g$ is related to $g_0$ as

\begin{equation}\label{G_Renormalization}
\frac{1}{g^2} = \frac{1}{g_0^2} \mp \frac{22 C_2}{3\varepsilon (4\pi)^3} + \mbox{higher orders},
\end{equation}

\noindent
which exactly agrees with the relation obtained in (the revised version of) Ref. \cite{Ivanov:2005qf}
and in Ref. \cite{Casarin:2019aqw}. It is more convenient to rewrite it in terms
of $\alpha \equiv g^2/4\pi$,

\begin{equation}
\frac{1}{\alpha} = \frac{1}{\alpha_0} \mp \frac{22C_2}{3\varepsilon(4\pi)^2} + \mbox{higher orders}.
\end{equation}

\noindent
This relation implies that the one-loop $\beta$-function is given by the expression

\begin{equation}
\beta(\alpha) =  \mp \frac{11\alpha^2 C_2}{24\pi^2} + \mbox{higher orders}.
\end{equation}

\noindent
Thus, the lower sign corresponds to the Landau zero,
while the upper sign corresponds to the asymptotic freedom. If one takes into account the difference in the notations, the
former choice was used in Ref. \cite{Ivanov:2005qf}, while the calculation of Ref. \cite{Casarin:2019aqw} corresponds to the latter
case.

\section{Conclusion}
\hspace*{\parindent}

In this paper we have considered the higher-derivative $6D,\, {\cal
N}=(1,0)$ supersymmetric  Yang-Mills theory in the harmonic
superspace formulation. The theory was quantized within the
background superfield  method which allows to preserve the manifest
gauge invariance and $6D$, ${\cal N}=(1,0)$ supersymmetry at all
stages of the quantum calculations. Using the supergraph technique
and the regularization by dimensional reduction we have calculated
the one-loop divergences of the quantum effective action. The
calculations were organized as follows. The divergences were firstly
found to the lowest order with respect to the background gauge
superfield. Then, using the manifest gauge invariance, the full
result for the divergent part of the one-loop effective action was
restored \footnote{As was already mentioned, an alternative to this
calculation is the computational procedure preserving the manifest
gauge invariance at all steps and based on the superfield
proper-time technique \cite{Buchbinder:2016url}. We plan to consider
this approach in the forthcoming paper.}. It was shown that the
divergences can be absorbed into the renormalization of the
dimensionless coupling constant $g_0$. The corresponding
$\beta$-function was computed and its sign was shown to be
determined by the sign of the initial classical action.

It is natural to expect that the theory will remain renormalizable after adding to the action (\ref{Action})
the action of $6D$, ${\cal N}=(1,0)$ gauge theory without higher derivatives involving
a dimensionful coupling constant $f_0$.
It would be interesting to study the renormalization properties of such a theory involving two coupling constants $g_0$ and $f_0$.
Another interesting problem is the construction of the $6D$, ${\cal N}=(1,1)$ higher-derivative supersymmetric
gauge theory and the study of its renormalization properties. For this purpose one should extend the theory considered here
by higher-derivative couplings with the hypermultiplet in the adjoint representation (e.g., along the line of Ref. \cite{Ivanov:2005kz})
and try to ensure the hidden ${\cal N}=(0, 1)$ supersymmetry in such an extended $6D$ system. These problems will be addressed elsewhere.

\section*{Acknowledgements}
\hspace*{\parindent}

I.L.B and E.A.I are grateful to Andrei Smilga for useful discussions in the course of this work. The work was
supported by Russian Science Foundation, grant No. 16-12-10306.

\appendix

\section*{Appendix}

\section{The second variation of the classical action}
\hspace*{\parindent}\label{Appendix_Variation}

For calculating the one-loop divergences we need to single out that part of the action
which is quadratic in the quantum gauge superfields. As is evident from Eq. (\ref{Splitting}),
to this end one should calculate the second variation of the action, then make the replacement

\begin{equation}\label{Replacement}
\delta V^{++}\ \to\ v^{++}; \qquad V^{++}\ \to\ \bm{V}^{++},
\end{equation}

\noindent
and finally multiply the result by $1/2$.

The first variation of the action (\ref{Action}) is given by the expression

\begin{equation}\label{First_Variation_Original}
\delta S = \pm \frac{1}{g_0^2}\,\mbox{tr} \int d\zeta^{(-4)} du\, F^{++} \delta F^{++} = \pm \frac{1}{g_0^2}\, \mbox{tr} \int d^{14}z\, du\, F^{++} \delta V^{--}.
\end{equation}

\noindent
To obtain the variation $\delta V^{--}$, note that, as a consequence of Eq. (\ref{Flatness}),
it should satisfy the relation

\begin{equation}\label{Equation_For_Variation}
\nabla^{++} \delta V^{--} = \nabla^{--} \delta V^{++},
\end{equation}

\noindent
where the covariant derivatives are defined as

\begin{equation}
\nabla^{++} \equiv D^{++} + i V^{++}, \qquad \nabla^{--} \equiv D^{--} + i V^{--}.
\end{equation}

\noindent
Note that, in contrast to Eqs. (\ref{Background_Covariant_Derivatives}), these equation
contain the superfields $V^{\pm\pm}$ (instead of $\bm{V}^{\pm\pm}$). The solution of Eq. (\ref{Equation_For_Variation}) satisfies the equation

\begin{equation}\label{Equation_For_Variation_V--}
\delta V^{--} = \frac{1}{2} \big(\nabla^{--}\big)^2 \delta V^{++} - \frac{1}{2} \nabla^{++} \nabla^{--} \delta V^{--}.
\end{equation}

\noindent
Substituting this expression into (\ref{First_Variation_Original}) and using (\ref{Equation_For_F}),
the first variation of the action can be presented in the form

\begin{equation}\label{First_Variation}
\delta S = \pm \frac{1}{2g_0^2}\, \mbox{tr} \int d^{14}z\, du\, F^{++} \big(\nabla^{--}\big)^2 \delta V^{++}.
\end{equation}

Next, we calculate the second variation

\begin{eqnarray}
&& \delta^2 S = \delta (\delta S) = \pm \frac{1}{2g_0^2}\, \mbox{tr} \int d^{14}z\, du\, \Big( \delta F^{++} \big(\nabla^{--}\big)^2 \delta V^{++} + i F^{++}\Big[\delta V^{--},\, \nabla^{--} \delta V^{++}\Big] \qquad \nonumber\\
&& + i F^{++} \nabla^{--} \Big[\delta V^{--},\, \delta V^{++}\Big] \Big).
\end{eqnarray}

\noindent
Substituting $\delta F^{++} = (D^+)^4 \delta V^{--}$ into this expression,  after some algebra we obtain

\begin{eqnarray}\label{Second_Variation_Original}
&& \delta^2 S = \pm \frac{1}{g_0^2}\, \mbox{tr} \int d^{14}z\, du\, \Big(\delta V^{--} \uB \delta V^{++}\nonumber\\
&&\qquad\qquad\qquad - \frac{i}{2} \delta V^{--} \Big[F^{++},\, \nabla^{--} \delta V^{++}\Big] + \frac{i}{2} \delta V^{--} \Big[\nabla^{--} F^{++},\, \delta V^{++}\Big] \Big),\qquad
\end{eqnarray}

\noindent
where

\begin{equation}
\uB\ \equiv \frac{1}{2} (D^+)^4 \big(\nabla^{--}\big)^2.
\end{equation}

To find the variation $\delta V^{--}$ we introduce the (non-analytic) bridge superfield $b$ related to $V^{\pm\pm}$
by the relations similar to (\ref{Background_Bridge}),

\begin{equation}\label{Bridge}
V^{++} = -i e^{i b} D^{++} e^{-i b};\qquad V^{--} = -i e^{i b} D^{--} e^{-i b}.
\end{equation}

\noindent
Then, taking into account that $\nabla^{\pm\pm} = e^{ib} D^{\pm\pm} e^{-ib}$, the solution of Eq. (\ref{Equation_For_Variation})
can be written as \cite{Buchbinder:2015wpa}

\begin{equation}\label{V--_Variation}
\delta V^{--}_1 = \int du_2\, \frac{1}{(u_1^+ u_2^+)^2}\, e^{ib_1} e^{-ib_2} \delta V^{++}_2 e^{i b_2} e^{-ib_1},
\end{equation}

\noindent
where the subscripts refer to the corresponding harmonic variables.

Substituting the expression (\ref{V--_Variation}) into Eq. (\ref{Second_Variation_Original})
we rewrite the second variation of the action (\ref{Action}) as

\begin{eqnarray}\label{Second_Variation_Of_The_Action}
&& \delta^2 S = \pm \frac{1}{g_0^2}\, \mbox{tr} \int d^{14}z\, du_1\, du_2\, \frac{1}{(u_1^+ u_2^+)^2}\,  e^{ib_1} e^{-ib_2} \delta V^{++}_2 e^{ib_2} e^{-ib_1} \qquad\nonumber\\
&&\qquad\qquad\qquad \times \Big( \uB_1 \delta V^{++}_1 - \frac{i}{2} \Big[F^{++}_1,\, \nabla^{--}_1 \delta V^{++}_1\Big] + \frac{i}{2} \Big[\nabla^{--}_1 F^{++}_1,\, \delta V^{++}_1\Big] \Big).\qquad
\end{eqnarray}

\noindent
Making in this expression the replacement (\ref{Replacement}) and multiplying the result by $1/2$
we obtain that part of the action (\ref{Action}) which is quadratic in the quantum gauge superfields,

\begin{eqnarray}\label{Quadratic_Part_Of_The_Action}
&& S^{(2)} = \pm \frac{1}{2g_0^2}\, \mbox{tr} \int d^{14}z\, du_1\, du_2\, \frac{1}{(u_1^+ u_2^+)^2}\,  e^{i\bm{b}_1} e^{-i\bm{b}_2} v^{++}_2 e^{i\bm{b}_2} e^{-i\bm{b}_1} \qquad\nonumber\\
&&\qquad\qquad\qquad\quad \times \Big( \sB_1 v^{++}_1 - \frac{i}{2} \Big[\bm{F}^{++}_1,\, \bm{\nabla}^{--}_1 v^{++}_1\Big] + \frac{i}{2} \Big[\bm{\nabla}^{--}_1 \bm{F}^{++}_1,\, v^{++}_1\Big] \Big).\qquad
\end{eqnarray}

To obtain the analogous part of the {\it total} action, we should add the gauge-fixing action (\ref{Gauge-Fixing_Action})
to this expression. Using the identity

\begin{eqnarray}
&& \big[ \sB, \bm{\nabla}^{++}\big] v^{++} = \frac{1}{2} (D^+)^4 \left[\big(\bm{\nabla}^{--}\big)^2, \bm{\nabla}^{++}\right] v^{++} = - \frac{1}{2} (D^+)^4 \Big(D^0\, \bm{\nabla}^{--} + \bm{\nabla}^{--} D^0\Big) v^{++} \qquad\nonumber\\
&& = - (D^+)^4 \bm{\nabla}^{--} v^{++} = -i[\bm{F}^{++}, v^{++}]
\end{eqnarray}

\noindent
after integrating by parts with respect to the harmonic derivatives the gauge-fixing term (\ref{Gauge-Fixing_Action})
can be rewritten as

\begin{eqnarray}\label{S_GF_Transformed}
&& S_{\mbox{\scriptsize gf}} = \mp \frac{1}{2 g_0^2\xi_0}\, \mbox{tr} \int d^{14}z\, du_1\, du_2\,
e^{i\bm{b}_1} e^{-i\bm{b}_2} v_{2}^{++}\, e^{i\bm{b}_2} e^{- i\bm{b}_1} \Big\{ D_1^{++} D_2^{++}
\Big(\frac{(u_1^- u_2^-)}{(u_1^+ u_2^+)^3}\Big)\, \sB_1 v_{1}^{++} \qquad \nonumber\\
&& + i D_2^{++} \Big(\frac{(u_1^- u_2^-)}{(u_1^+ u_2^+)^3}\Big)\, [\bm{F}_1^{++}, v_1^{++}] \Big\}.\qquad
\end{eqnarray}

\noindent
Then, with the help of the identities

\begin{eqnarray}
&& D_1^{++} D_2^{++}\Big(\frac{(u_1^- u_2^-)}{(u_1^+ u_2^+)^3}\Big) = \frac{1}{(u_1^+ u_2^+)^2}
- \frac{1}{2} \big(D_1^{--}\big)^2 \delta^{(2,-2)}(u_1,u_2)\,, \quad\\
&& D_2^{++} \Big(\frac{(u_1^- u_2^-)}{(u_1^+ u_2^+)^3}\Big) = \frac{(u_1^- u_2^+)}{(u_1^+ u_2^+)^3} = - \frac{1}{2} D_1^{--} \Big(\frac{1}{(u_1^+ u_2^+)^2}\Big)
\end{eqnarray}

\noindent
it is convenient to rearrange the expression (\ref{S_GF_Transformed}) to the form

\begin{eqnarray}\label{S_GF_Final}
&& S_{\mbox{\scriptsize gf}} = \pm \frac{1}{2 g_0^2\xi_0}\, \mbox{tr} \int d\zeta^{(-4)}\, du\, v^{++} \sB{}^2 v^{++} \mp \frac{1}{2 g_0^2\xi_0}\, \mbox{tr} \int d^{14}z\, du_1\, du_2\, \frac{1}{(u_1^+ u_2^+)^2}\,  \qquad\nonumber\\
&& \times\, e^{i\bm{b}_1} e^{-i\bm{b}_2} v_{2}^{++}\, e^{i\bm{b}_2} e^{- i\bm{b}_1} \Big\{ \sB_1 v_{1}^{++} + \frac{i}{2}\, \bm{\nabla}^{--}_1 [\bm{F}_1^{++}, v_1^{++}] \Big\}.\qquad
\end{eqnarray}

The quadratic in the quantum gauge superfield part of the total action  is obtained as a sum of
the expressions (\ref{Quadratic_Part_Of_The_Action}) and (\ref{S_GF_Final}),

\begin{eqnarray}\label{Quadratic_Part_Final}
&&\hspace*{-7mm} S^{(2)} + S_{\mbox{\scriptsize gf}} = \pm \frac{1}{2 g_0^2\xi_0}\, \mbox{tr} \int d\zeta^{(-4)}\, du\, v^{++} \sB{}^2 v^{++} \pm \frac{1}{2g_0^2}\, \mbox{tr} \int d^{14}z\, du_1\, du_2\, \frac{1}{(u_1^+ u_2^+)^2}\, e^{i\bm{b}_1} e^{-i\bm{b}_2} v_{2}^{++}\, \nonumber\\
&&\hspace*{-7mm} \times\,  e^{i\bm{b}_2} e^{- i\bm{b}_1} \Big\{ \Big(1-\frac{1}{\xi_0}\Big) \sB_1 v_{1}^{++} + \frac{i}{2}\Big(1-\frac{1}{\xi_0}\Big)\, [\bm{\nabla}^{--}_1 \bm{F}_1^{++}, v_1^{++}] - \frac{i}{2}\Big(1+\frac{1}{\xi_0}\Big)\, [ \bm{F}_1^{++}, \bm{\nabla}^{--}_1 v_1^{++}] \Big\}.\nonumber\\
\end{eqnarray}

\noindent
This expression is drastically simplified in the Feynman gauge $\xi_0=1$.
In this gauge Eq. (\ref{Quadratic_Part_Final}) is reduced to the expression (\ref{Second_Variation}).

\section{Divergences of the supergraphs with a gauge loop}
\hspace*{\parindent}\label{Appendix_Divergences}

Let us calculate the one-loop divergences coming from the first two supergraphs
presented in Fig. \ref{Figure_Diagrams}. Both these supergraphs contain a loop of the quantum gauge
superfield, so that the vertices can be found from the expression (\ref{Second_Variation}). In particular,
the triple vertex present in the supergraph (1) can be written as

\begin{eqnarray}\label{Triple_Vertex}
&& \mp \frac{1}{4g_0^2} f^{ABC} \int d^{14}z\, du\, \partial^2 v^{++A} \Big[\bm{V}_{\mbox{\scriptsize linear}}^{--B} D^{--} v^{++C} + D^{--} \Big(\bm{V}_{\mbox{\scriptsize linear}}^{--B} v^{++C} \Big)\Big]\quad\nonumber\\
&& \pm \frac{1}{4g_0^2} f^{ABC} \int d^{14}z\, du_1\, du_2\, \frac{1}{(u_1^+ u_2^+)^2}\, v_2^{++A} \bm{F}_{\mbox{\scriptsize linear},1}^{++B} D_1^{--} v_1^{++C} \equiv \mbox{Ver}_1 + \mbox{Ver}_2,\quad
\end{eqnarray}

\noindent
where the subscript ``linear'' means that it is enough to consider only the part linear in the background gauge superfield $\bm{V}^{++}$,

\begin{equation}
\bm{V}_{\mbox{\scriptsize linear},1}^{--A} = \int du_2\, \frac{1}{(u_1^+ u_2^+)^2}\, \bm{V}^{++A}_2; \qquad \bm{F}_{\mbox{\scriptsize linear},1}^{++A} = \int du_2\, \frac{1}{(u_1^+ u_2^+)^2}\, (D_1^+)^4 \bm{V}^{++A}_2.
\end{equation}

\noindent
We see that the vertex (\ref{Triple_Vertex}) can naturally be divided into the two parts, $\mbox{Ver}_1$ and $\mbox{Ver}_2$.
The first one is composed of the terms which include $\bm{V}_{\mbox{\scriptsize linear}}^{--B}$, while the second one contains
the term with $\bm{F}_{\mbox{\scriptsize linear}}^{++B}$. Therefore, the supergraph (1) in Fig. \ref{Figure_Diagrams}
splits into three subgraphs, namely, $\mbox{Ver}_1 - \mbox{Ver}_1$, $\mbox{Ver}_2 - \mbox{Ver}_2$, and $\mbox{Ver}_1 - \mbox{Ver}_2$.
The subgraph containing two vertices $\mbox{Ver}_1$ is very similar to an analogous supergraph
calculated in Ref. \cite{Buchbinder:2017ozh} and vanishes due to the presence of the factors

\begin{equation}\label{Vanishing Expressions}
\big(D_{1}^{--}\big)^2 (u_1^+ u_2^+)^4 \Big|_{u_1=u_2} = 0\qquad \mbox{or}\qquad D_{1}^{--} D_2^{--} (u_1^+ u_2^+)^4 \Big|_{u_1=u_2} = 0.
\end{equation}

\noindent
The subgraph $\mbox{Ver}_2 - \mbox{Ver}_2$ is proportional to

\begin{equation}
\int \frac{d^6k}{(2\pi)^6}\frac{1}{k^4 (k+p)^4}
\end{equation}

\noindent
and is, therefore, finite. This implies that this subgraph does not contribute to the divergent part
of the one-loop effective action and can be omitted.

Thus, we see that the only non-trivial contribution of the supergraph (1) to the one-loop divergences
can appear from the (logarithmically divergent) subgraph $\mbox{Ver}_1 - \mbox{Ver}_2$.
Applying Feynman rules, we find that the considered contribution to the effective action is given by the expression

\begin{eqnarray}\label{Nontrivial_Superdiagram}
&&\hspace*{-9mm} \mbox{Ver}_1 - \mbox{Ver}_2 = - \frac{i C_2}{4} \int d^{14}z_1 d^{14}z_2\, du_1\, du_2\, du_3\, \frac{1}{(u_2^+ u_3^+)^2}\, \bm{V}_{\mbox{\scriptsize linear}}^{--A}(z_1,u_1)\, \bm{F}_{\mbox{\scriptsize linear}}^{++A}(z_2,u_2) \nonumber\\
&&\hspace*{-9mm} \times \bigg\{D^{--}_2\Big[\frac{1}{\partial^4} (D^+_2)^4 \delta^{14}(z_1-z_2)\, \delta^{(2,-2)}(u_1,u_2)\Big]\, D^{--}_1 \Big[\frac{1}{\partial^2}(D^+_3)^4 \delta^{14}(z_1-z_2)\, \delta^{(2,-2)}(u_1,u_3)\Big] \nonumber\\
&&\hspace*{-9mm} - D^{--}_1 D^{--}_2 \Big[\frac{1}{\partial^4} (D^+_2)^4\delta^{14}(z_1-z_2)\, \delta^{(2,-2)}(u_1,u_2) \Big]\, \frac{1}{\partial^2} (D^+_3)^4 \delta^{14}(z_1-z_2)\, \delta^{(2,-2)}(u_1,u_3) \nonumber\\
&&\hspace*{-9mm} + D^{--}_2\Big[\frac{1}{\partial^2} (D^+_2)^4 \delta^{14}(z_1-z_2)\, \delta^{(2,-2)}(u_1,u_2)\Big]\, D^{--}_1 \Big[\frac{1}{\partial^4}(D^+_3)^4 \delta^{14}(z_1-z_2)\, \delta^{(2,-2)}(u_1,u_3)\Big] \nonumber\\
&&\hspace*{-9mm} - D^{--}_1 D^{--}_2 \Big[\frac{1}{\partial^2} (D^+_2)^4\delta^{14}(z_1-z_2)\, \delta^{(2,-2)}(u_1,u_2) \Big]\, \frac{1}{\partial^4} (D^+_3)^4 \delta^{14}(z_1-z_2)\, \delta^{(2,-2)}(u_1,u_3) \bigg\}.
\end{eqnarray}

\noindent
Note that, without a loss of generality, we can assume that all spinor derivatives act on the point $z_1$.

The $\delta$-functions $\delta^{14}(z_1-z_2)$ include $\delta^8(\theta_1-\theta_2)$, and the product of two Grassmannian
$\delta$-functions does not vanish only in the situation when at least 8 spinor covariant derivatives act on them.
Taking into account this property and using the identity

\begin{equation}
\delta^8(\theta_1-\theta_2)\, (D^+_1)^4 (D^+_2)^4 \delta^8(\theta_1-\theta_2) = (u_1^+ u_2^+)^4 \delta^8(\theta_1-\theta_2)
\end{equation}

\noindent
we can do one of the integrals over $d^8\theta$. Then, in the momentum representation,
Eq. (\ref{Nontrivial_Superdiagram}) can be brought in the form

\begin{eqnarray}
&& \frac{i C_2}{2} \int d^8\theta\, du_1\, du_2\, du_3\, \int \frac{d^6p}{(2\pi)^6}\, \bm{V}_{\mbox{\scriptsize linear}}^{--A}(-p,\theta,u_1)\, \bm{F}_{\mbox{\scriptsize linear}}^{++A}(p,\theta,u_2)\, \frac{1}{(u_2^+ u_3^+)^2} \qquad \nonumber\\
&& \times  \int \frac{d^6k}{(2\pi)^6}\, \frac{1}{k^4 (k+p)^2} \bigg\{D^{--}_2\Big[(u_2^+ u_3^+)^4 \delta^{(2,-2)}(u_1,u_2)\Big]\, D^{--}_1 \delta^{(2,-2)}(u_1,u_3)\nonumber\\
&& - D^{--}_1 D^{--}_2 \Big[(u_2^+ u_3^+)^4 \delta^{(2,-2)}(u_1,u_2) \Big]\, \delta^{(2,-2)}(u_1,u_3) \bigg\}. \vphantom{\frac{d^6k}{(2\pi)^6}}
\end{eqnarray}

\noindent
After integrating by parts with respect to the derivative $D^{--}_1$ in the last term, this expression
can be rewritten as

\begin{eqnarray}
&& \frac{i C_2}{2} \int d^8\theta\, du_1\, du_2\, du_3\, \int \frac{d^6p}{(2\pi)^6}\,\frac{d^6k}{(2\pi)^6}\, \frac{1}{k^4 (k+p)^2}\, \bm{F}_{\mbox{\scriptsize linear}}^{++A}(p,\theta,u_2)\, \bigg\{\frac{2}{(u_2^+ u_3^+)^2} \qquad \nonumber\\
&& \times\, \bm{V}_{\mbox{\scriptsize linear}}^{--A}(-p,\theta,u_1)\, D^{--}_2\Big[(u_2^+ u_3^+)^4 \delta^{(2,-2)}(u_1,u_2)\Big]\, D^{--}_1 \delta^{(2,-2)}(u_1,u_3) + (u_2^+ u_3^+)^2 \qquad\vphantom{\frac{d^6k}{(2\pi)^6}}\nonumber\\
&& \times\, D^{--}_1 \bm{V}_{\mbox{\scriptsize linear}}^{--A}(-p,\theta,u_1)\, D^{--}_2 \delta^{(2,-2)}(u_1,u_2)\, \delta^{(2,-2)}(u_1,u_3) \bigg\}, \vphantom{\frac{d^6k}{(2\pi)^6}}
\end{eqnarray}

\noindent
where we took into account that

\begin{equation}
(u_2^+ u_3^+)\, \delta^{(2,-2)}(u_1,u_2)\, \delta^{(2,-2)}(u_1,u_3)\ \to\ 0.
\end{equation}

\noindent
Once again, integrating by parts with respect to the derivative $D^{--}_2$ and doing the integral over $du_1$
with the help of $\delta^{(2,-2)}(u_1,u_2)$, we obtain

\begin{eqnarray}
&& \mbox{Ver}_1 - \mbox{Ver}_2 = \frac{i C_2}{2} \int d^8\theta\, du_2\, du_3\, \int \frac{d^6p}{(2\pi)^6}\, \frac{d^6k}{(2\pi)^6}\, \frac{1}{k^4 (k+p)^2} \qquad \nonumber\\
&& \times   \bigg\{4 \bm{V}_{\mbox{\scriptsize linear}}^{--A}(-p,\theta,u_2)\, \bm{F}_{\mbox{\scriptsize linear}}^{++A}(p,\theta,u_2)\, (u_2^- u_3^+)\, (u_2^+ u_3^+)\, D^{--}_2 \delta^{(2,-2)}(u_2,u_3) \qquad\nonumber\\
&& - 2 \bm{V}_{\mbox{\scriptsize linear}}^{--A}(-p,\theta,u_2)\, D^{--}_2 \bm{F}_{\mbox{\scriptsize linear}}^{++A}(p,\theta,u_2)\, (u_2^+ u_3^+)^2\, D^{--}_2 \delta^{(2,-2)}(u_2,u_3) \vphantom{\frac{d^6k}{(2\pi)^6}}\nonumber\\
&& - 2 D^{--}_2 \bm{V}_{\mbox{\scriptsize linear}}^{--A}(-p,\theta,u_2)\, \bm{F}_{\mbox{\scriptsize linear}}^{++A}(p,\theta,u_2)\, (u_2^- u_3^+)\, (u_2^+ u_3^+)\, \delta^{(2,-2)}(u_2,u_3)
\vphantom{\frac{d^6k}{(2\pi)^6}}\nonumber\\
&& - D^{--}_2 \bm{V}_{\mbox{\scriptsize linear}}^{--A}(-p,\theta,u_2)\, D^{--}_2 \bm{F}_{\mbox{\scriptsize linear}}^{++A}(p,\theta,u_2)\, (u_2^+ u_3^+)^2 \, \delta^{(2,-2)}(u_2,u_3) \bigg\}. \vphantom{\frac{d^6k}{(2\pi)^6}}
\end{eqnarray}

\noindent
Two last terms in this expression evidently vanish because they contain

\begin{equation}\label{Vanishing_Product}
(u_2^+ u_3^+)\, \delta^{(2,-2)}(u_2,u_3)\ \to\ 0.
\end{equation}

\noindent
Integrating by parts with respect to the derivative $D^{--}_2$ in the remaining terms we see that the only expression which does not contain the vanishing product (\ref{Vanishing_Product}) is

\begin{eqnarray}
&& - 2 i C_2 \int d^8\theta\, du_2\, du_3\, \int \frac{d^6p}{(2\pi)^6}\, \frac{d^6k}{(2\pi)^6}\, \frac{1}{k^4 (k+p)^2} \qquad \nonumber\\
&&\qquad\qquad\qquad\qquad \times  \bm{V}_{\mbox{\scriptsize linear}}^{--A}(-p,\theta,u_2)\, \bm{F}_{\mbox{\scriptsize linear}}^{++A}(p,\theta,u_2)\, (u_2^- u_3^+)^2\, \delta^{(2,-2)}(u_2,u_3). \qquad
\end{eqnarray}

\noindent
After calculating the harmonic integrals, this expression takes the form

\begin{eqnarray}
- 2 i C_2 \int d^8\theta\, du\, \int \frac{d^6p}{(2\pi)^6}\, \frac{d^6k}{(2\pi)^6}\, \frac{1}{k^4 (k+p)^2}  \bm{V}_{\mbox{\scriptsize linear}}^{--A}(-p,\theta,u)\, \bm{F}_{\mbox{\scriptsize linear}}^{++A}(p,\theta,u).
\end{eqnarray}

The momentum integral in this expression is calculated in the Euclidean space after the standard Wick rotation
 with the help of dimensional reduction. Clearly, we are interested in its divergent part only,

\begin{equation}
\int \frac{d^6k}{(2\pi)^6} \frac{1}{k^4 (k+p)^2}\ \to\ -i\int \frac{d^DK}{(2\pi)^D} \frac{1}{K^6} + \mbox{finite terms} = -\frac{i}{\varepsilon (4\pi)^3}  + \mbox{finite terms},
\end{equation}

\noindent
where the capital letter $K$ denotes the Euclidean loop momentum.

Thus, in the coordinate representation we obtain the following divergent contribution coming from the supergraph (1) in Fig. \ref{Figure_Diagrams}:

\begin{equation}\label{One_Loop_Divergence_Quadratic}
\Big[\mbox{diagram}\, (1)\Big]_\infty = - \frac{2 C_2}{\varepsilon(4\pi)^3} \int d^{14}z\,du\, \bm{V}_{\mbox{\scriptsize linear}}^{--A}\, \bm{F}_{\mbox{\scriptsize linear}}^{++A} = -\frac{4 C_2}{\varepsilon(4\pi)^3} \mbox{tr} \int d\zeta^{(-4)} du\, \big(\bm{F}_{\mbox{\scriptsize linear}}^{++}\big)^2.
\end{equation}

Now, let us calculate the contribution of the tadpole supergraph (2) in Fig. (\ref{Figure_Diagrams}).
The relevant vertex is also obtained from the expression (\ref{Second_Variation}) and is composed out of the terms
quadratic in the background gauge superfield $\bm{V}^{++}$ and in the background bridge $\bm{b}$,

\begin{eqnarray}\label{Quadric_Vertex}
&& \pm \frac{1}{2g_0^2} \mbox{tr} \int d^{14}z\, du\, \Big\{i \partial^2 v^{++} [\bm{V}^{--}_{\mbox{\scriptsize quadratic}}, D^{--} v^{++}] + i \partial^2 v^{++}  D^{--} [\bm{V^{--}}_{\mbox{\scriptsize quadratic}}, v^{++}]\nonumber\\
&& - \partial^2 v^{++} [\bm{V}_{\mbox{\scriptsize linear}}^{--}, [\bm{V}^{--}_{\mbox{\scriptsize linear}}, v^{++} ]] - \frac{1}{4} \Big([\bm{V}^{--}_{\mbox{\scriptsize linear}}, D^{--} v^{++}] + D^{--} [\bm{V^{--}}_{\mbox{\scriptsize linear}}, v^{++}]\Big) (D^+)^4 \quad\nonumber\\
&& \times \Big([\bm{V}^{--}_{\mbox{\scriptsize linear}}, D^{--} v^{++}] + D^{--}[\bm{V^{--}}_{\mbox{\scriptsize linear}}, v^{++}]\Big) \Big\} \pm \frac{1}{2g_0^2} \mbox{tr} \int d^{14}z\,du_1\, du_2\, \frac{1}{(u_1^+ u_2^+)^2}\,\nonumber\\
&& \times \Big\{ [ \bm{b}_1 - \bm{b}_2,\,  v_2^{++}]\, [\bm{F}_{\mbox{\scriptsize linear},1}^{++}, D^{--} v^{++}_1] -i v_2^{++} [\bm{F}_{\mbox{\scriptsize quadratic},1}^{++}, D^{--} v^{++}_1]\Big\}, \vphantom{\frac{1}{2}}
\end{eqnarray}

\noindent
where $\bm{V}^{--}_{\mbox{\scriptsize quadratic}}$ and $\bm{F}_{\mbox{\scriptsize quadratic}}^{++}$ are
those parts of $\bm{V}^{--}$ and $\bm{F}^{++}$ which are quadratic in $\bm{V}^{++}$. Constructing
the corresponding expression for the contribution of the supergraph (2), we see that all terms
in it vanish either because the number of spinor derivatives acting on the anticommuting $\delta$-function
at the coincident arguments is less than 8 (for all terms without $(D^+)^4$ in Eq. (\ref{Quadric_Vertex})),
or because of the presence of $(u_1^+ u_2^+)\Big|_{u_1=u_2} = 0$ (for the terms containing $(D^+)^4$).
Therefore, the supergraph (2) in Fig. \ref{Figure_Diagrams} does not contribute to the divergent part
of the one-loop effective action.

Thus, the one-loop divergences coming from the supergraphs (1) and (2) in Fig. \ref{Figure_Diagrams}
are given by Eq. (\ref{One_Loop_Divergence_Quadratic}). However, the gauge superfield $\bm{V}^{++}$
is dimensionless, so that the one-loop divergences are also present in supergraphs with an arbitrary number of
the external gauge legs. The general result for the divergent part of the one-loop effective action
can be restored by resorting to the manifest gauge invariance of the effective action $\Gamma[\bm{V}^{++}]$.
\footnote{Note that the in the case of using the background field method one-loop contributions generated
by the supergraphs with a gauge loop and by the supergraphs with a ghost loop are gauge invariant separately.}
The gauge invariant expression which in the lowest approximation yields Eq. (\ref{One_Loop_Divergence_Quadratic})
is evidently given just by Eq. (\ref{Gauge_Contribution}).


\begin{thebibliography}{100}

\bibitem{Fradkin:1982kf}
  E.~S.~Fradkin and A.~A.~Tseytlin,
  ``Quantum Properties of Higher Dimensional and Dimensionally Reduced Supersymmetric Theories,''
  Nucl.\ Phys.\ B {\bf 227} (1983) 252.

\bibitem{Marcus:1983bd}
  N.~Marcus and A.~Sagnotti,
  ``A Test of Finiteness Predictions for Supersymmetric Theories,''
  Phys.\ Lett.\  {\bf 135B} (1984) 85.

\bibitem{Marcus:1984ei}
  N.~Marcus and A.~Sagnotti,
  Nucl.\ Phys.\ B {\bf 256} (1985) 77.

\bibitem{Howe:1983jm}
  P.~S.~Howe and K.~S.~Stelle,
  ``Ultraviolet Divergences in Higher Dimensional Supersymmetric {Yang-Mills} Theories,''
  Phys.\ Lett.\  {\bf 137B} (1984) 175.

\bibitem{Howe:2002ui}
  P.~S.~Howe and K.~S.~Stelle,
  ``Supersymmetry counterterms revisited,''
  Phys.\ Lett.\ B {\bf 554} (2003) 190,
  {\tt arXiv:hep-th/0211279}.

\bibitem{Bossard:2009sy}
  G.~Bossard, P.~S.~Howe and K.~S.~Stelle,
  ``The Ultra-violet question in maximally supersymmetric field theories,''
  Gen.\ Rel.\ Grav.\  {\bf 41} (2009) 919,
  {\tt arXiv:0901.4661 [hep-th]}.

\bibitem{Bossard:2009mn}
  G.~Bossard, P.~S.~Howe and K.~S.~Stelle,
  ``A Note on the UV behaviour of maximally supersymmetric Yang-Mills theories,''
  Phys.\ Lett.\ B {\bf 682} (2009) 137,
  {\tt arXiv:0908.3883 [hep-th]}.

\bibitem{Bork:2015zaa}
  L.~V.~Bork, D.~I.~Kazakov, M.~V.~Kompaniets, D.~M.~Tolkachev and D.~E.~Vlasenko,
  ``Divergences in maximal supersymmetric Yang-Mills theories in diverse dimensions,''
  JHEP {\bf 1511} (2015) 059,
  {\tt arXiv:1508.05570 [hep-th]}.

\bibitem{Pais:1950za}
  A.~Pais and G.~E.~Uhlenbeck,
  ``On Field theories with nonlocalized action,''
  Phys.\ Rev.\  {\bf 79} (1950) 145.

\bibitem{Starobinsky:1980te}
  A.~A.~Starobinsky,
  ``A New Type of Isotropic Cosmological Models Without Singularity,''
  Phys.\ Lett.\  {\bf 91B} (1980) 99
   [Adv.\ Ser.\ Astrophys.\ Cosmol.\  {\bf 3} (1987) 130].

\bibitem{Hawking:1985gh}
  S.~W.~Hawking,
  ``Who's Afraid Of (Higher Derivative) Ghosts?,''
   in Quantum Field Theory and Quantum Statistics, edited by
I.~A.~Batalin, C.~J.~Isham, G.~A.~Vilkovisky, VOL. 2, 129-139, 1986,
Adam Hilger Publishing, 1986.

\bibitem{Stelle:1976gc}
  K.~S.~Stelle,
  ``Renormalization of Higher Derivative Quantum Gravity,''
  Phys.\ Rev.\ D {\bf 16} (1977) 953.

\bibitem{Fradkin:1981iu}
  E.~S.~Fradkin and A.~A.~Tseytlin,
  ``Renormalizable asymptotically free quantum theory of gravity,''
 Nucl.\ Phys.\ B {\bf 201} (1982) 469.

\bibitem{Buchbinder:1992rb}
  I.~L.~Buchbinder, S.~D.~Odintsov and I.~L.~Shapiro,
  ``Effective Action in Quantum Gravity,''
  Bristol and Philadelphia, UK: IOP (1992) 413 p.

\bibitem{Grinstein:2008qq}
  B.~Grinstein and D.~O'Connell,
  ``One-Loop Renormalization of Lee-Wick Gauge Theory,''
  Phys.\ Rev.\ D {\bf 78} (2008) 105005,
  {\tt arXiv:0801.4034 [hep-ph]}.

\bibitem{Beccaria:2015ypa}
  M.~Beccaria and A.~A.~Tseytlin,
  ``Conformal anomaly c-coefficients of superconformal 6d theories,''
  JHEP {\bf 1601} (2016) 001,
  {\tt [arXiv:1510.02685 [hep-th]]}.

\bibitem{Osborn:2016bev}
  H.~Osborn and A.~Stergiou,
  ``C$_{T}$ for non-unitary CFTs in higher dimensions,''
  JHEP {\bf 1606} (2016) 079,
  {\tt [arXiv:1603.07307 [hep-th]]}.

\bibitem{Kuzenko:2017jdy}
  S.~M.~Kuzenko, J.~Novak and S.~Theisen,
  ``New superconformal multiplets and higher derivative invariants in six dimensions,''
  Nucl.\ Phys.\ B {\bf 925} (2017) 348,
  {\tt [arXiv:1707.04445 [hep-th]]}.

\bibitem{Castellanos:2018dub}
  A.~R.~R.~Castellanos, F.~Sobreira, I.~L.~Shapiro and A.~A.~Starobinsky,
  ``On higher derivative corrections to the $R+R^2$ inflationary model,''
  JCAP {\bf 1812} (2018) 007,
 {\tt  arXiv:1810.07787 [gr-qc]}.

\bibitem{Anselmi:2020tqo}
  D.~Anselmi,
  ``The quest for purely virtual quanta: fakeons versus Feynman-Wheeler particles,''
  JHEP {\bf 2003} (2020) 142,
  {\tt arXiv:2001.01942 [hep-th]}.

\bibitem{Smilga:2016dpe}
  A.~Smilga,
  ``Ultraviolet divergences in non-renormalizable supersymmetric theories,''
  Phys.\ Part.\ Nucl.\ Lett.\  {\bf 14} (2017) no.2,  245,
  {\tt  arXiv:1603.06811 [hep-th]}.

\bibitem{Casarin:2017xez}
  L.~Casarin,  ``On higher-derivative gauge theories,''
  {\tt arXiv:1710.08021 [hep-th]}.

\bibitem{Slavnov:1971aw}
  A.~A.~Slavnov,
  ``Invariant regularization of nonlinear chiral theories,''
  Nucl.\ Phys.\ B {\bf 31} (1971) 301.

\bibitem{Slavnov:1972sq}
  A.~A.~Slavnov,
  ``Invariant regularization of gauge theories,''
  Theor.Math.Phys. {\bf 13} (1972) 1064
   [Teor.\ Mat.\ Fiz.\  {\bf 13} (1972) 174].

\bibitem{Krivoshchekov:1978xg}
  V.~K.~Krivoshchekov,
  ``Invariant Regularizations for Supersymmetric Gauge Theories,''
  Theor.\ Math.\ Phys.\ {\bf 36} (1978) 745
 [Teor.\ Mat.\ Fiz.\  {\bf 36} (1978) 291].

\bibitem{West:1985jx}
  P.~C.~West,
  ``Higher Derivative Regulation of Supersymmetric Theories,''
  Nucl.\ Phys.\ B {\bf 268} (1986) 113.

\bibitem{Buchbinder:2014wra}
  I.~L.~Buchbinder and K.~V.~Stepanyantz,
  ``The higher derivative regularization and quantum corrections in N=2 supersymmetric theories,''
  Nucl.\ Phys.\ B {\bf 883} (2014) 20,
  {\tt arXiv:1402.5309 [hep-th]}.

\bibitem{Buchbinder:2015eva}
  I.~L.~Buchbinder, N.~G.~Pletnev and K.~V.~Stepanyantz,
  ``Manifestly N=2 supersymmetric regularization for N=2 supersymmetric field theories,''
  Phys.\ Lett.\ B {\bf 751} (2015) 434,
  {\tt arXiv:1509.08055 [hep-th]}.

\bibitem{Gates:1983nr}
  S.~J.~Gates, M.~T.~Grisaru, M.~Rocek and W.~Siegel,
  ``Superspace Or One Thousand and One Lessons in Supersymmetry,''
  Front.\ Phys.\  {\bf 58} (1983) 1.
  [hep-th/0108200].

\bibitem{West:1990tg}
  P.~C.~West,
  ``Introduction to supersymmetry and supergravity,''
  Singapore, Singapore: World Scientific (1990) 425 p.

\bibitem{Buchbinder:1998qv}
  I.~L.~Buchbinder and S.~M.~Kuzenko,
  ``Ideas and methods of supersymmetry and supergravity: Or a walk through superspace,''
  Bristol, UK: IOP (1998) 656 p.

\bibitem{Stepanyantz:2019lyo}
  K.~Stepanyantz,
  ``The higher covariant derivative regularization
  as a tool for revealing the structure of quantum
  corrections in supersymmetric gauge theories,''
  {\tt arXiv:1910.03242 [hep-th]}.

\bibitem{Stepanyantz:2019lfm}
  K.~V.~Stepanyantz,
  ``The NSVZ $\beta$-function for theories regularized by higher covariant derivatives: the all-loop sum of matter and ghost singularities,''
  JHEP {\bf 2001} (2020) 192,
  {\tt arXiv:1912.12589 [hep-th]}.

\bibitem{Ivanov:2005qf}
  E.~A.~Ivanov, A.~V.~Smilga and B.~M.~Zupnik,
  ``Renormalizable supersymmetric gauge theory in six dimensions,''
  Nucl.\ Phys.\ B {\bf 726} (2005) 131,
  {\tt arXive:hep-th/0505082}.

\bibitem{Casarin:2019aqw}
  L.~Casarin and A.~A.~Tseytlin,
  ``One-loop $\beta$-functions in 4-derivative gauge theory in 6 dimensions,''
  JHEP {\bf 1908} (2019) 159,
  {\tt arXiv:1907.02501 [hep-th]}.

\bibitem{Buchbinder:2020ovf}
  I.~L.~Buchbinder, E.~A.~Ivanov, B.~S.~Merzlikin and K.~V.~Stepanyantz,
  ``The renormalization structure of $6D$, ${\cal N}=(1,0)$ supersymmetric higher-derivative gauge theory,''
  {\tt arXiv:2007.02843 [hep-th]}.

\bibitem{Smilga:2006ax}
  A.~V.~Smilga,
  ``Chiral anomalies in higher-derivative supersymmetric 6D theories,''
  Phys.\ Lett.\ B {\bf 647} (2007) 298,
  {\tt arXiv:hep-th/0606139}.


\bibitem{Howe:1985ar}
  P.~S.~Howe, K.~S.~Stelle and P.~C.~West,
  ``N=1, d = 6 harmonic superspace'',
  Class.\ Quant.\ Grav.\  {\bf 2} (1985) 815-821.

\bibitem{Zupnik:1986da}
  B.~M.~Zupnik,
  ``Six-dimensional Supergauge Theories in the Harmonic Superspace,''
  Sov.\ J.\ Nucl.\ Phys.\  {\bf 44} (1986) 512
   [Yad.\ Fiz.\  {\bf 44} (1986) 794].

\bibitem{Bossard:2015dva}
  G.~Bossard, E.~Ivanov and A.~Smilga,
  ``Ultraviolet behavior of 6D supersymmetric Yang-Mills theories and harmonic superspace'',
  JHEP {\bf 1512} (2015) 085,
  {\tt arXiv:1509.08027 [hep-th]}.

\bibitem{Galperin:1985ec}
  A.~Galperin, E.~Ivanov, V.~Ogievetsky and E.~Sokatchev,
  ``Harmonic Superspace: Key To N=2 Supersymmetry Theories,''
  JETP Lett.\  {\bf 40} (1984) 912
   [Pisma Zh.\ Eksp.\ Teor.\ Fiz.\  {\bf 40} (1984) 155].

\bibitem{Galperin:1984av}
  A.~Galperin, E.~Ivanov, S.~Kalitzin, V.~Ogievetsky and E.~Sokatchev,
  ``Unconstrained N=2 Matter, Yang-Mills and Supergravity Theories in Harmonic Superspace,''
  Class.\ Quant.\ Grav.\  {\bf 1} (1984) 469
   Erratum: [Class.\ Quant.\ Grav.\  {\bf 2} (1985) 127].

\bibitem{Galperin:2001uw}
  A.~S.~Galperin, E.~A.~Ivanov, V.~I.~Ogievetsky and E.~S.~Sokatchev,
  ``Harmonic superspace'',
  Cambridge, UK: Univ. Pr. (2001) 306 p.

\bibitem{Buchbinder:2016gmc}
  I.~L.~Buchbinder, E.~A.~Ivanov, B.~S.~Merzlikin and K.~V.~Stepanyantz,
  ``One-loop divergences in the $6D, \mathcal N = (1,0)$ abelian gauge theory,''
  Phys.\ Lett.\ B {\bf 763} (2016) 375,
  {\tt arXiv:1609.00975 [hep-th]}.

\bibitem{Buchbinder:2016url}
  I.~L.~Buchbinder, E.~A.~Ivanov, B.~S.~Merzlikin and K.~V.~Stepanyantz,
  ``One-loop divergences in 6D, $ \mathcal{N} $ = (1, 0) SYM theory,''
  JHEP {\bf 1701} (2017) 128,
  {\tt arXiv:1612.03190 [hep-th]}.

\bibitem{Buchbinder:2017ozh}
  I.~L.~Buchbinder, E.~A.~Ivanov, B.~S.~Merzlikin and K.~V.~Stepanyantz,
  ``Supergraph analysis of the one-loop divergences in $6D$, ${\cal N} = (1,0)$ and ${\cal N} = (1,1)$ gauge theories,''
  Nucl.\ Phys.\ B {\bf 921} (2017) 127,
  {\tt arXiv:1704.02530 [hep-th]}.

\bibitem{Buchbinder:2017gbs}
  I.~L.~Buchbinder, E.~A.~Ivanov, B.~S.~Merzlikin and K.~V.~Stepanyantz,
  ``On the two-loop divergences of the 2-point hypermultiplet supergraphs for $6D$, ${\cal N} = (1,1)$ SYM theory,''
  Phys.\ Lett.\ B {\bf 778} (2018) 252,
  {\tt arXiv:1711.11514 [hep-th]}.

\bibitem{Buchbinder:2017xjb}
  I.~L.~Buchbinder, E.~A.~Ivanov and B.~S.~Merzlikin,
  ``Leading low-energy effective action in $6D$, ${\cal N}=(1,1)$ SYM theory,''
  JHEP {\bf 1809} (2018) 039,
  {\tt arXiv:1711.03302 [hep-th]}

\bibitem{Buchbinder:2018bhs}
  I.~L.~Buchbinder, E.~Ivanov, B.~Merzlikin and K.~Stepanyantz,
  ``Harmonic Superspace Approach to the Effective Action in Six-Dimensional Supersymmetric Gauge Theories,''
  Symmetry {\bf 11} (2019) no.1,  68,
  {\tt arXiv:1812.02681 [hep-th]}.

\bibitem{Buchbinder:2018lbd}
  I.~L.~Buchbinder, E.~A.~Ivanov, B.~S.~Merzlikin and K.~V.~Stepanyantz,
  ``Gauge dependence of the one-loop divergences in $6D$, ${\cal N} = (1,0)$ abelian theory,''
  Nucl.\ Phys.\ B {\bf 936} (2018) 638,
  {\tt arXiv:1808.08446 [hep-th]}.

\bibitem{Zupnik:1987vm}
  B.~M.~Zupnik,
  ``The Action of the Supersymmetric $N=2$ Gauge Theory in Harmonic Superspace,''
  Phys.\ Lett.\ B {\bf 183} (1987) 175.

\bibitem{DeWitt:1965jb}
  B.~S.~DeWitt,
  ``Dynamical theory of groups and fields,''
  Gordon and Breach, New York, 1965.

\bibitem{Abbott:1980hw}
  L.~F.~Abbott,
  ``The Background Field Method Beyond One Loop,''
  Nucl.\ Phys.\ B {\bf 185} (1981) 189.

\bibitem{Abbott:1981ke}
  L.~F.~Abbott,
  ``Introduction to the Background Field Method,''
  Acta Phys.\ Polon.\ B {\bf 13} (1982) 33.

\bibitem{Buchbinder:1997ya}
  I.~L.~Buchbinder, E.~I.~Buchbinder, S.~M.~Kuzenko and B.~A.~Ovrut,
  ``The background field method for N=2 superYang-Mills theories in harmonic superspace'',
  Phys.\ Lett.\ B {\bf 417} (1998) 61-71,
  {\tt arXiv:hep-th/9704214}.

\bibitem{Buchbinder:2001wy}
  E.~I.~Buchbinder, B.~A.~Ovrut, I.~L.~Buchbinder, E.~A.~Ivanov and S.~M.~Kuzenko,
  ``Low-energy effective action in N = 2 supersymmetric field theories,''
  Phys.\ Part.\ Nucl.\  {\bf 32} (2001) 641
   [Fiz.\ Elem.\ Chast.\ Atom.\ Yadra {\bf 32} (2001) 1222].

\bibitem{Ivanov:2005kz}
  E.~A.~Ivanov and A.~V.~Smilga,
  ``Conformal properties of hypermultiplet actions in six dimensions,''
  Phys.\ Lett.\ B {\bf 637} (2006) 374,
  {\tt [hep-th/0510273]}.

\bibitem{Buchbinder:2015wpa}
  I.~L.~Buchbinder and N.~G.~Pletnev,
  ``Effective actions in $ \mathcal{N}=1 $ , D5 supersymmetric gauge theories: harmonic superspace approach,''
  JHEP {\bf 1511} (2015) 130,
  {\tt arXiv:1510.02563 [hep-th]}.

\end{thebibliography}
\end{document}